\documentclass[a4paper]{article}
\usepackage{amsmath}
\usepackage{hyperref}
\usepackage[english]{babel}
\pdfoutput=1 

\hypersetup{pdftex, colorlinks=true, linkcolor=blue, citecolor=blue, filecolor=blue, urlcolor=blue, pdftitle=Artificial Cosmogenesis: A New Kind of Cosmology, pdfauthor=Clement Vidal, pdfsubject=, pdfkeywords={artificial cosmogenesis, cosmic evolution, computational cosmology, digital physics, Drake equation, Cosmic Evolution Equation, robustness, fine-tuning, multiverse}}
\usepackage[pdftex]{graphicx}

\begin{document}
\date{}
\title{Artificial Cosmogenesis: \\ A New Kind of Cosmology}
\author{\textbf{Cl\'ement Vidal}}
\maketitle 
\begin{center} Center Leo Apostel\\
Global Brain Institute\\
Evolution, Complexity and Cognition research group\\
Vrije Universiteit Brussel (Free University of Brussels)\\
Krijgskundestraat 33, 1160 Brussels, Belgium\\
Phone +32-2-640 67 37 {\textbar} Fax +32-2-6440744\\ 
{\ }\href{http://clement.vidal.philosophons.com/}{http://clement.vidal.philosophons.com}\\
\href{mailto:clement.vidal@philosophons.com}{clement.vidal@philosophons.com} \\ 
\end{center}
\bigskip

\noindent in Zenil, H. 2012 (ed) \textit{Irreducibility and Computational Equivalence: Wolfram Science 10 Years After the Publication of A New Kind of Science} (to appear, v1.0).

\bigskip

\textbf{Abstract}: This paper introduces foundations for a new kind of
cosmology. We advocate that computer simulations are needed to address
two key cosmological issues. First, the \textit{robustness} of the
emergence of complexity, which boils down to ask: ``what would remain
the same if the tape of the universe were replayed?'' Second, the much
debated \textit{fine-tuning} issue, which requires to answer the
question: \ ``are complex universes rare or common in the space of
possible universes?'' We argue that computer simulations are
indispensable tools to address those two issues scientifically. We first
discuss definitions of \textit{possible universes} and of
\textit{possible cosmic outcomes} -- such as atoms, stars, life or
intelligence. This leads us to introduce a generalized Drake-like
equation, the \textit{Cosmic Evolution Equation}. It is a modular and
conceptual framework to define research agendas in computational
cosmology. We outline some studies of alternative complex
universes. However, such studies are still in their infancy, and they
can be fruitfully developed within a new kind of cosmology, heavily
supported by computer simulations, \textit{Artificial Cosmogenesis}. The
appendix [section \ref{maps}] provides argumentative maps of the paper's
main thesis.

\bigskip

\pagebreak

\tableofcontents
\bigskip
{\raggedleft\itshape
What I cannot create I do not understand
\par}
{\raggedleft\
\ \ On Richard Feynman's blackboard \\ at time of death in 1988, as reported in \cite{hawking}
\par}

\section{Introduction}

I am fond of both computer science and cosmology. However, the methods,
concepts and tools used in those two disciplines are very different. Is
it possible to unite this dual passion? This essay outlines foundations
for such a new kind of cosmology, \textit{Artificial Cosmogenesis}.

Broadly speaking, we can distinguish three kinds of science:
\textit{deterministic}, \textit{probabilistic} and
\textit{computational}. Deterministic science can roughly be
characterized by the science Newton practiced. He used physical laws and
initial conditions to predict the future and explain the past. The
predictions are of an amazing accuracy and the tools used are
mathematical equations which are relatively easy to solve. Because of
its successes, it is often implicitly considered the typical model of
hard science.

However, when there are too many particles in a system, their sheer
number and interactions make the Newtonian approach weak. In fact, even
with only three gravitational bodies the Newtonian theory of gravitation
fails to make practically useful predictions. The main insight of the
founders of statistical physics was to average out the interactions of
particles to derive \textit{statistical} laws of behavior, such as the
laws of thermodynamics or quantum mechanics.

In recent years, Laurent Nottale generalized this statistical
predictability to all scales in nature, by unifying relativity theories
with microphysics (see e.g. \cite{nottalea,nottaleb,nottalec}). This
\textit{scale relativity} theory constitutes a revolution in progress in
the domain of theoretical physics, since its leads to fundamental
theoretical results as well as highly precise and validated predictions
(see also \cite{vidalintro}, p96-97).

But what if our statistical methods also fail or are absent? What if we
do not know any way to predict the behavior of a very complex system? An
even more general approach is needed. This can be done in a
computational view of nature, by theorizing and experimenting with
algorithms (see e.g.  \cite{zuse, mindbendingpaper}). The field of
Artificial Life constitutes a remarkable application of this view, when
it attempts to decipher the most general laws of life, and then to
implement and experiment with them in computers. Stephen Wolfram
\cite{wolframbook} argued at length how important this new kind of
science based on computer simulations is. He advocated a wide
exploration of simple programs, to study their behavior and
properties. He argued that such a new approach is unavoidable if we want
to understand complex dynamics. As a matter of fact, the study of
complex dynamical systems will in most cases \textit{not} be predictable
with simple equations. Wolfram \cite{wolframa,wolframbook} further
conjectured that most systems in nature are \textit{computationally
  irreducible}. This means that to study complex systems, there is no
shorter way than to run step by step the model, and study how it behaves
(see also \cite{zwirn} for a general formal definition of irreducible
computation). Such a kind of science can still make predictions because
simulations can be run faster than reality. Studying complex systems,
equations won't help, simulations will.

Of course, when possible, it is best to aim for absolute and precise
predictions such as in Newtonian science. When this fails, statistical
laws are the second best option. But most real and complex systems may
not be predictable in these two ways. A broader general computational
exploration promises to be the way to understand the rise and evolution
of complexity.

My aim in this paper is to propose a computational approach to progress
on two arduous cosmological issues. First, the \textit{robustness of the
  emergence of complexity} in our universe; second, the question of how
\textit{fine-tuned} our universe is.

The question of the robustness of the emergence of complexity can simply
be illustrated by a thought experiment. \textit{What would remain the
  same if we would replay the tape of the universe?} To address this
issue, we introduce the \textit{Cosmic Evolution Equation} (CEE). It is
a modular conceptual framework to discuss possible universes, possible
cosmic outcomes, the robustness of the universe and fine-tuning. To
define it, we build on Drake's \cite{drake} equation in the Search for
Extraterrestrial Intelligence (SETI) and on the thoughtful discussion of
possible universes by Ellis, Kirchner and Stoeger \cite{elliskirchstoe}.

The fine-tuning issue is much debated and intricate. The problem is that
if we vary one by one a number of parameters, both in cosmological and
standard particle models, no life or no complexity of any sort emerges
(see e.g. \cite{leslie, rees, daviesb}). The issue is mined with logical
and probabilistic fallacies (e.g. \cite{mcgrew, colyvan}) as well as
physical fallacies (see e.g. \cite{vidaledu,vidalphd,stengerbook}). It
is also commonly confused with other related issues such as free
parameters, parameter sensitivity, metaphysical issues, anthropic
principles, observational selection effects, teleology and God's
existence \cite{vidalphd}.

Additionally, different discussions of fine-tuning focus on very
different \textit{cosmic outcomes}. We see fine-tuning discussions
regarding the dimensionality of space \cite{rees}, the production of
carbon atoms in stars \cite{hoyle}, the existence of long-lived stars
\cite{adams}; the number of black holes \cite{smolin}; biochemistry
\cite{barrow}; but also complexity of any sort \cite{ellisa}. A key
question to clarify the issue is thus to explicitly ask:
\textit{fine-tuning for what?} Which cosmic outcome are we interested
in? In particular, we will see that most fine-tuning \ arguments are
poor, since they vary parameters one by one, which is a fallacy
resulting in exploring only 0.00000000000000456 \% of the parameter
space under consideration!

To remedy this situation, we generalize the CEE. The Drake equation
estimates the number of communicative intelligent civilizations in our
galaxy. By extension, one application of the generalized CEE is to
estimate the likelihood of our particular universe in the space of
possible universes. In other words, if Drake's equation allows to
estimate the probability of life existing ``somewhere in the galaxy'';
one application of the CEE is to estimate the more general probability
of life existing ``\textit{anywhere} in the space of possible
universes''. \textit{Artificial Cosmogenesis} -- ACosm for short -- is
the study of alternative cosmic evolutions and allows in principle to
assess how fine-tuned our universe is.

We first discuss the issues of possible universes and possible cosmic
outcomes (sections \ref{section2} and \ref{section3}). Then we introduce
the CEE to discuss the robustness issue (section \ref{section4}) and
generalize the CEE to address the fine-tuning issue (sections
\ref{section5} to \ref{section6}). By bridging the gap between computer
science and cosmology, I hope this framework will fruitfully pave the
way for resolving these two fundamental cosmological issues.

\section{Possible Universes}\label{section2}

What are the possible universes? How can we describe the space of
possible universes? These questions raise enormous logical,
metaphysical, philosophical, and scientific problems. Although possible
universes or possible worlds have been discussed throughout in the
history of philosophy (see e.g. \cite{leibniz, lewis}, see also
\cite{dick} for a wider historical perspective), our aim here is to
formulate the issue of possible universes so that it can progressively
leave the realm of metaphysics and enter operational science.

We now follow Ellis', Kirchner's and Stoeger's \cite{elliskirchstoe}
definition of the class of all possible universes. Let $M$ be a
structural and dynamical space of all possible universes $m$. Each
universe $m$ is described by a set of states $s$ in a state space
$S$. Each universe $m$ is characterized by a set $P$ of distinguishing
parameters $p$, which are coordinates on $S$. Such parameters will be
logical, physical or dynamical. How will they dynamically evolve? The
three authors elaborate:
\begin{quote}
Each universe $m$ will evolve from its initial state to some final state
according to the dynamics operative, with some or all of its parameters
varying as it does so. The course of this evolution of states will be
represented by a path in the state space $S$, depending on the
parametrisation of $S$. Thus, each such path (in degenerate cases a
point) is a representation of one of the universes $m$ in $M$. The
coordinates in $S$ will be directly related to the parameters specifying
members of $M$.
\end{quote}
In such a procedure, we face a first major issue:
\begin{quote}
\textbf{Possibility space issue:} \textit{What delimits the set of
  possibilities? What is the meta-law or meta-cause which determines
  $M$?}
\end{quote}
As the three authors argue, we can't avoid the meta-law issue, because
otherwise we have no basis to set up a consistent description of $M$. We
need to have a logic which describes $M$. There are other difficult
issues related to identifying which different representations represent
the same universe models -- \textit{the equivalence problem --} and the
problem of dealing with an \textit{infinite space of possible
  universes}. I refer the reader to the three authors' paper for more in
depth discussions of these issues.

More directly related to the fine-tuning issue is the remark of
Jean-Philippe Uzan that ``the larger the possibility space considered,
the more fine-tuned the actual universe appears to be'' (in
\cite{elliskirchstoe}, p923). Indeed, we can easily increase the
unlikelihood of our universe simply by allowing the parameter space to
grow. You could ask for example, did you explore if universes with 42
dimensions generate life? Do we really want to capture the radical idea
of ``all that can happen, happens''? There is much variation in the
space of possibility we want to delimit. Ellis (\cite{ellisb}, p1261)
distinguishes four levels of variation, \textit{weak},
\textit{moderate}, \textit{strong} and \textit{extreme}:
\begin{itemize}
\item ``\textit{Weak variation}: e.g. only the values of the constants
  of physics are allowed to vary? This is an interesting exercise but is
  certainly not an implementation of the idea `all that can happen,
  happens'. It is an extremely constrained set of variations.
\item \textit{Moderate variation}: different symmetry groups, or numbers
  of dimensions, etc. We might for example consider the possibility
  landscapes of string theory \cite{freivogel} as realistic indications
  of what may rule multiverses \cite{freivogel, susskind,carr}. But that
  is very far indeed from `all that is possible', for that should
  certainly include spacetimes not ruled by string theory.
\item \textit{Strong variation}: different numbers and kinds of forces,
  universes without quantum theory or in which relativity is untrue
  (e.g. there is an aether), some in which string theory is a good
  theory for quantum gravity and others where it is not, some with quite
  different bases for the laws of physics (e.g. no variational
  principles).
\item \textit{Extreme variation}: universes where physics is not well
  described by mathematics; with different logic; universes ruled by
  local deities; allowing magic as in the Harry Potter series of books;
  with no laws of physics at all? Without even mathematics or logic?''
\end{itemize}
We indeed need to make a choice between theoretical physics and
magic... or anything in between. \\ 

Do we need to assume an actual multiverse? No we do not. To study the
fine-tuning issue, we need only \textit{possible} or \textit{virtual}
universes, not actually realized ones. This interpretation still allows
us to use the vast multiverse literature to define and explore possible
universes, without making strong and problematic ontological claims
regarding their actual existence.

\section{Possible Cosmic Outcomes}\label{section3}

Once we settle on a framework to define possible universes, a second
major issue is to specify the parameters which differentiate possible
universes:
\begin{quote}
\textbf{Cosmic outcomes issue}: \textit{What are the cosmic outcomes?
  What are the milestones of cosmic evolution? What parameters
  differentiate possible universes? How do we find those parameters? }
\end{quote}
As the three authors mention, the values of the parameters may not be
known initially. They may emerge out of \textit{transitions} from one
regime to another. For example, sociologists do not explore alternative
sociological structures by varying the mass of elementary
particles. They start from different, less fundamental parameters, such
as the influence of population density, the climate or the
media. \textit{The challenge to understand complexity transitions in
  cosmic evolution is of upmost importance and difficulty}. For example,
how did atoms emerge out of the big bang era? How did planets form out
of stars and stardust? How did life originate out of molecules? How did
consciousness emerge from biological organisms? Etc.

The ideal of reducing such parameters is a major goal of science. The
objective is to build a consistent theory and narrative of cosmic
evolution, which explains a maximum of cosmic outcomes with a minimum of
parameters. Scientific progress is achieved when new theories capture
previously free and unexplained parameters (see e.g. \cite{vidaledu} for
an illustration in physics). We could now extend this attitude to
attempt a reduction of other higher parameters (such as life) to
fundamental physics and cosmic parameters. However, since we are still
very far from such a feat, in our description of possible universes we
must assume explicitly higher parameters. Typically, when researchers
tackle the issue of the origin of life, they don't start from big bang
nucleosynthesis, but they assume the existence of molecules.

Ellis, Kirchner and Stoeger categorize the parameters from the most
basic ones to the most complex ones. They distinguish different
categories of parameters $p_j$, with $j = 1 - 2$ describing basic
physics; $j = 3 - 5$ describing cosmology and a category of parameters
$j = 6 - 7$ related to the emergence of life and higher complexity.

Each category $p_j$ is composed of different parameters $i$. For
example, $p_1(i)$ are basic physics parameters, such that the
fine-structure constant; masses, charges and spins of particles, as well
as other dimensionless parameters. I refer the reader to the detailed
description of the parameters given by the three authors.

However, in each parameter category I would like to add explicitly some
random, chance or noise parameters. For example, these could include for
$j = 1 - 5$ quantum effects in the early universe; or nonlinear chaotic
dynamics which might trigger catastrophic events, such as meteorites
impacting planets for $j = 7$. This would certainly complicate the
dynamics, but would also make it much more realistic. A dynamical
argument can even be advanced that such random events might be essential
to the open-ended growth of complexity. An illustration can be found in
engineering with the heuristic of \textit{simulated annealing}. It
starts by adding important noise into the system, and then gradually
reduces it. The purpose of the noise is to shake the system to reach a
maximally stable configuration.

Now, how do we decide which cosmic outcomes to keep, and which ones to
leave out? At first, we can aim at including a maximum of
parameters. Then, we would progressively reduce the number of
parameters, as we get better and better insights on how they emerge from
more fundamental principles and theories; i.e. from previous
parameters. Robert Aunger (\cite{aungera}, p1142-1144) did compile from
many authors a list of more than 100 different cosmic outcomes. This is
the most comprehensive review I am aware of, ranging from the big bang,
the formation of atoms, stars, solar systems, life, DNA,
multicellularity, sexual reproduction, fishes, to mammals, agriculture,
modern science and space exploration.

However, we can already anticipate a fallacy lurking when considering a
large list of cosmic outcomes. Similarly to Uzan's remark for the space
of possible universes, we can note that the more cosmic outcomes we
have, the more unlikely they will seem. The extreme case is to consider
one single object as a cosmic outcome. For example, in intelligent
design discussions, they consider a complex object (like a living
organism or an airplane) and try to assess the likelihood that it arose
by chance. Of course this will be very unlikely! Additionally, as
Dawkins \cite{dawkins} argues, natural selection would still constitute
a much better candidate explanation than design. A scientist will look
for possible mechanisms, theories, which can explain the emergence of
complexity. The \textit{a posteriori} probability of a single object
isolated from its evolutionary or human context is of weak scientific
interest.

To avoid such an error, we need to advance \textit{theoretical reasons}
to select certain cosmic outcomes and not others. This is rarely
attempted. Most authors propose an arbitrary list without strong
theoretical justification. Ellis, Kirchner and Stoeger did not justify
their choice of distinguishing parameters; although it is clear that
they included a lot of cosmological parameters necessary for their
subsequent study of alternative universes with different geometries.

The most promising avenue of research is to focus on thermodynamics (see
e.g. \cite{salthe}). Indeed, all systems need to process energy, which
is therefore a universal concept, applicable from the beginning of the
universe to our energy hungry technological society. Robert Aunger
\cite{aungera,aungerb} built on a thermodynamical theory to select
cosmic outcomes, \textit{non-equilibrium steady-state transitions}. Each
transition involves first an energy innovation, then a structural
adjustment and finally a new control mechanism. He thus constructed a
consistent selection of cosmic outcomes and evolutionary transitions.

Which cosmic outcomes are contingent and evolutionary? Which ones are
necessary and developmental? Are there attractors in the dynamic of
cosmic evolutionary development? To answer these issues, we need to
explore the \textit{robustness} of the emergence of complexity. Stated
otherwise, if we would re-run the tape of the universe, would galaxies,
stars, biology and technology arise again and again? The straightforward
way to answer those question, in parallel to a theoretical rationale
like Aunger's, is indeed to re-run the tape of the universe. Let us now
examine how we can conceptualize and do that.

\section{Robustness in Cosmic Evolution}\label{section4}

{\raggedleft 
\itshape {what would remain the same if the tape of life
    were replayed? \par}} 
{\raggedleft\ Stephen Jay Gould \cite{gould} \par}

{\raggedleft
\itshape {what would remain the same if the tape of the
    universe were replayed?  \par}}
{\raggedleft\
Paraphrasing Gould's question to the universe \cite{vidalacosm}
\par} 

\bigskip

Answering this latter question, Paul Davies (\cite{daviesa}, p317) wrote
that if ``the universe were re-run a second time, there would be no
solar system, no Earth and no people. But the emergence of life and
consciousness somewhere and somewhen in the cosmos is, I believe,
assured by the underlying laws of nature.'' Those claims, as Davies
acknowledges, are only informed intuitions. How can we test this
intuition or different ones scientifically? This is the issue of the
\textit{robustness of the emergence of complexity in cosmic evolution}.

A first analysis of the tape metaphor shows its limits. Indeed, if the
tape and its player were perfect, we should get exactly the same results
when re-running the tape. So, the thought experiment would be
trivial. Yet if our universe self-constructs, one question is whether
small fluctuations, chance events, noise or random perturbations would
lead to slightly different outcomes, or very different ones. This makes
the issue of robustness in cosmic evolution highly stimulating.

It is very hard to tackle because it is linked to a great weakness of
cosmology as a science: it has only one object of study, our unique
universe. More precisely, we can distinguish two fundamental limitations
that Ellis (\cite{ellisb}, 1216) pointed out:

\begin{quote}
\textbf{Thesis A1: The universe itself cannot be subjected to physical
  experimentation}. \textit{We cannot re-run the universe with the same
  or altered conditions to see what would happen if they were different,
  so we cannot carry out scientific experiments on the universe itself.}
Furthermore,
\end{quote}

\begin{quote}
\textbf{Thesis A2: The universe cannot be observationally compared with
  other universes}. \textit{We cannot compare the universe with any
  similar object, nor can we test our hypotheses about it by
  observations determining statistical properties of a known class of
  physically existing universes.}
\end{quote}

Our thesis is that it is possible to address those limitations and the
issue of robustness by running computer simulations of our universe. It
is important to note that if we replay the tape of \textit{our}
universe, we don't aim to actually explore the full space of possible
universes. Here, we only aim to assess the robustness of the emergence
of the different cosmic outcomes. We thus vary \textit{only}
nondeterministic dynamical parameters we discussed above (quantum
mechanical effects, random perturbations, nonlinear chaotic dynamics,
etc.). An open question is also how we vary the random parameters. How
often? How strong is the variation? Various distributions can be tested,
from Gaussian distributions, where most random variations are of an
average strength, few are weak or strong; to power-law distributions,
where there are few very strong variations, some medium variations, and
most of the time weak random variations.

Because of the inclusion of such parameters, it makes sense to re-run
the same universe simulation. By running a multitude of times the
simulation, it will be possible to make statistics on the emergence of
complexity. An even more straightforward way to make such statistics
would be to drastically intensify astrobiology -- the search for
extraterrestrials. If or when we will find extraterrestrials, we would
be able to progressively study the ``natural re-runs'' of
complexity, elsewhere than on Earth. Additionally, searching for extraterrestrials more complex
than us would force us to break with the implicit anthropocentric
assumption that life and humans on Earth are the highest development in
cosmic evolution. This invites us to speculate on the existence of
higher cosmic outcomes, and this opens the way to test our theories of
the general evolution of cosmic complexity (see e.g. \cite{bradbury,
  vidalbh} for modern views on the search for advanced
extraterrestrials).

An example of ambitious simulations of \textit{our }universe are the
Millennium run simulations \cite{springel,boylan,guo}. The authors
studied the formation, evolution and clustering of galaxies and quasars
within the standard (or concordance) model of cosmology. Although they
did not run the same simulation in its full complexity many times, the
volume space explored is large enough to extract meaningful statistical
properties on the evolution of the distribution of matter.

Replaying the tape of our entire universe is still a much more ambitious project, which at present remains unrealistic. We should remain aware that our current models and their associated free parameters are most likely not the ultimate ones. Of course, new theories need to be developed to know what the key parameters of our universe are. In the meantime, a way to progress is to break down the issue into smaller solvable problems. For example, if we want to tackle the robustness up to the emergence of intelligent life, we can write a generalized Drake equation (\cite{elliskirchstoe}, p925) that we call the \textit{Cosmic Evolution Equation}:
\begin{equation*}      
\textit{N}_{life} (m^{* })  = \textit{ N}_g \cdot  \textit{ N}_S \cdot  \textit{ f}_S \cdot \textit{ f}_p \cdot  \textit{ n}_e \cdot \textit{ f}_l \cdot \textit{ f}_i  
\end{equation*}
where $\textit{N}_{life} (\textit{m}^{*})$ is the number of planets with
intelligent life in our particular universe $\textit{m}^{*}$; and
\begin{itemize}
\item $\textit{N}_g$ is the number of galaxies in the model
\item $\textit{N}_S$ is the average number of stars per galaxy
\item $\textit{f}_S$ is the fraction of stars suitable for life
\item $\textit{f}_p$ is the fraction of such stars with planetary systems
\item $\textit{n}_e$ is the mean number of planets which are suitable habitats for life
\item $\textit{f}_l$ is the fraction of planets on which life originates
\item $\textit{f}_i$ is the fraction of life bearing planets with intelligent life.
\end{itemize}
There are many implicit assumptions in such a framework, for example that life-supporting stars will be Sun-like; or that life starts necessarily on planets and not on more exotic places. We also implicitly assume that the parameters are independent. To deal with dependent parameters, one would need to introduce a Bayesian probability framework. Additionally, we may have clear definitions of what stars or galaxies are, but the issues of defining higher cosmic outcomes such as life or intelligence remain of huge scientific debate.

The factors $\textit{N}_g$ and $\textit{N}_S$ can nowadays be estimated,
while the recent explosion of exoplanets discoveries is allowing us to
estimate more and more precisely the factors $\textit{f}_S \cdot
\textit{f}_p \cdot \textit{n}_e$. However, huge uncertainties remain
regarding the last two factors $\textit{f}_l \cdot \textit{f}_i$.

The main interest of such a framework --whether we consider these seven
factors to be most relevant or others-- is that we can in a first
approximation estimate the factors independently. Additionally,
\textit{the more we progress in our knowledge of the universe, the
  larger the distance between factors we can assess}. For example,
assessing the number of planets with intelligent life knowing only the
number of galaxies seems very hard. But shorter distances between
factors are easier to assess. For example, Miller's \cite{miller} famous
experiment tells us that the probability to have amino acids out of a
primordial soup and some energy source is high. Which is indeed an
important insight to evaluate $\textit{n}_e \cdot \textit{f}_l$.

Let us now imagine that we run multiple times a model of our entire
universe $m^*$. We would be able to interpret the results of the
multiple runs of the simulation as a set of \textit{virtual}
universes. We would end up with a distribution function $f(m^*)$
combining the probability distributions obtained for each factor.
However, we need to further specify a \textit{possibility space}, which
in this case is $M^*$ resulting from the variation of random parameters
only; and a mathematical measure $\pi^*$ on $M^*$. Such a virtual ensemble of
simulated universes $V$ would thus be defined as:
\begin{equation*}      
V=\{M^*, \pi^*, f(m^*)\}
\end{equation*}
The number of planets with intelligent life would then be:
\begin{equation*}       
\textit{N}_{life} (m^{* })  = \int \textit{ N}_g \cdot  \textit{ N}_S \cdot  \textit{ f}_S \cdot \textit{ f}_p \cdot  \textit{ n}_e \cdot \textit{ f}_l \cdot \textit{ f}_i \cdot \pi^*
\end{equation*}
Note that the integral is necessary to normalize the result according to
the measure $\pi^*$ and distribution function $f(m^*)$.  There are
important and subtle issues to make this normalization sound and
possible (see again\cite{elliskirchstoe}).

Let us give some more concrete possible results such simulation studies
would bring. We might conclude that our universe is robust for
galaxy-formation, i.e. most simulation runs lead to galaxy formation.
But still, it might turn out that our universe is not robust for
intelligent life, i.e. most simulations \textit{do not} lead to the
emergence of intelligent life.

We can now take a fresh look on our question: are cosmic outcomes
necessary or contingent? We can define a cosmic outcome as
\textit{necessary} if it appears again and again as we re-run the same
universe simulation, as \textit{contingent} otherwise. For example, let
us take the DNA code in biology: is it necessary that there is a unique
DNA code for terrestrial or extraterrestrial biology? In a similar
fashion, in economy, is it a necessity in civilizational development
that monetary systems converge to a common currency?

We can also compare the cosmic outcome selections. On the one hand we
would have the ones resulting from ``simulation experiments'' (see
e.g. \cite{kleijnen} for a discussion); and on the other hand the
theoretical approaches (such as Aunger's). \textit{Simulation
  experiments} in cosmology can play the role that \textit{empirical
  experiments} play in other sciences. This approach can be called
``cosmology in silico'' or ``computational cosmology''. In fact, these
endeavors are already developing quickly, as illustrated by the Virgo
Consortium for Cosmological Supercomputer Simulations.

We have just begun to explore how robust the emergence of complexity in
our universe is. If we want to understand it better, we need to perform
computer simulations and use existing conceptual, mathematical and
statistical tools to design simulation experiments and to assess the
results.

However interesting and important this enterprise is, it does not tackle
the fine-tuning issue. Indeed, in studying the robustness of our
universe, we try to understand the emergence of complexity in
\textit{our universe}, whereas to address fine-tuning we must study the
place of our particular universe in \textit{the space of possible
  universes}.

\section{Artificial Cosmogenesis \\ or the study of alternative cosmic evolutions}\label{section5}

{\raggedleft\itshape {Now, we create a considerable problem. For we are
    tempted to make statements of comparative reference regarding the
    properties of our observable Universe with respect to the
    alternative universes we can imagine possessing different values of
    their foundamental constants. But there is only one Universe; where
    do we find the other possible universes against which to compare our
    own in order to decide how fortunate it is that all these remarkable
    coincidences that are necessary for our own evolution actually
    exist?  \par}} 
{\raggedleft\ Barrow and Tipler (\cite{barrowtipler},  p6) \par}

\bigskip

{\raggedleft\itshape {you might end up having a future subject which is
    ``comparative universality'' -- we have all these laws for the
    universe that cannot be eliminated as ours and you study them, you
    talk about them, you compare them, this could be a future
    subject. Students would be required to pass exams on their ten
    possible favorite universes ...  \par}} 
{\raggedleft\ Gregory Chaitin (\cite{chaitin}, p339) \par}

\bigskip

This first quote by Barrow and Tipler summarizes the core problem of
fine-tuning. The second quote by Chaitin illustrates a core idea towards
its resolution. With the robustness issue, we have focused on
\textit{our} universe. To assess in how far our universe is fine-tuned,
we must study the place of our universe in the \textit{space of possible
  universes}. We call this space the \textit{virtual multiverse}.

Fine-tuning arguments vary just one parameter, a fallacy which is nearly
\textit{always} committed. The underlying assumption is that parameters
are independent. As Stenger (\cite{stengerbook}, p70) remarks, this is
``both dubious and scientifically shoddy''. If the history of physics
learned us something is that phenomena which where thought to be widely
independent, turned out to have common underlying causes and
principles. For example, our common sense fails to see a connection
between the fall of an apple and the tides; magnetism and electricity;
and even less between space, time and the speed of light. But all these
phenomena have been unified thanks to physical theories.

Additionally, varying several parameters without care can lead to what
is known as the \textit{one-factor-at-a-time} (OAT) paradox in
sensitivity analysis. The problem with the OAT method is that it is
non-explorative. Let us see why. At first sight, it seems logical and
rigorous, since it varies factors one-at-a-time while keeping the others
constant. It seems consistent because the output from a change can be
attributed unambiguously to the change of one factor. It also never
detects non-influential factors as relevant. However, by construction,
this method is non-explorative, with exploration decreasing rapidly with
the number of factors. For a simple example, consider Figure 1, which
shows clearly that OAT explores only 5 points forming a cross, out of 9
points in total.

\begin{figure} 
\begin{center} 
\includegraphics[scale=0.6]{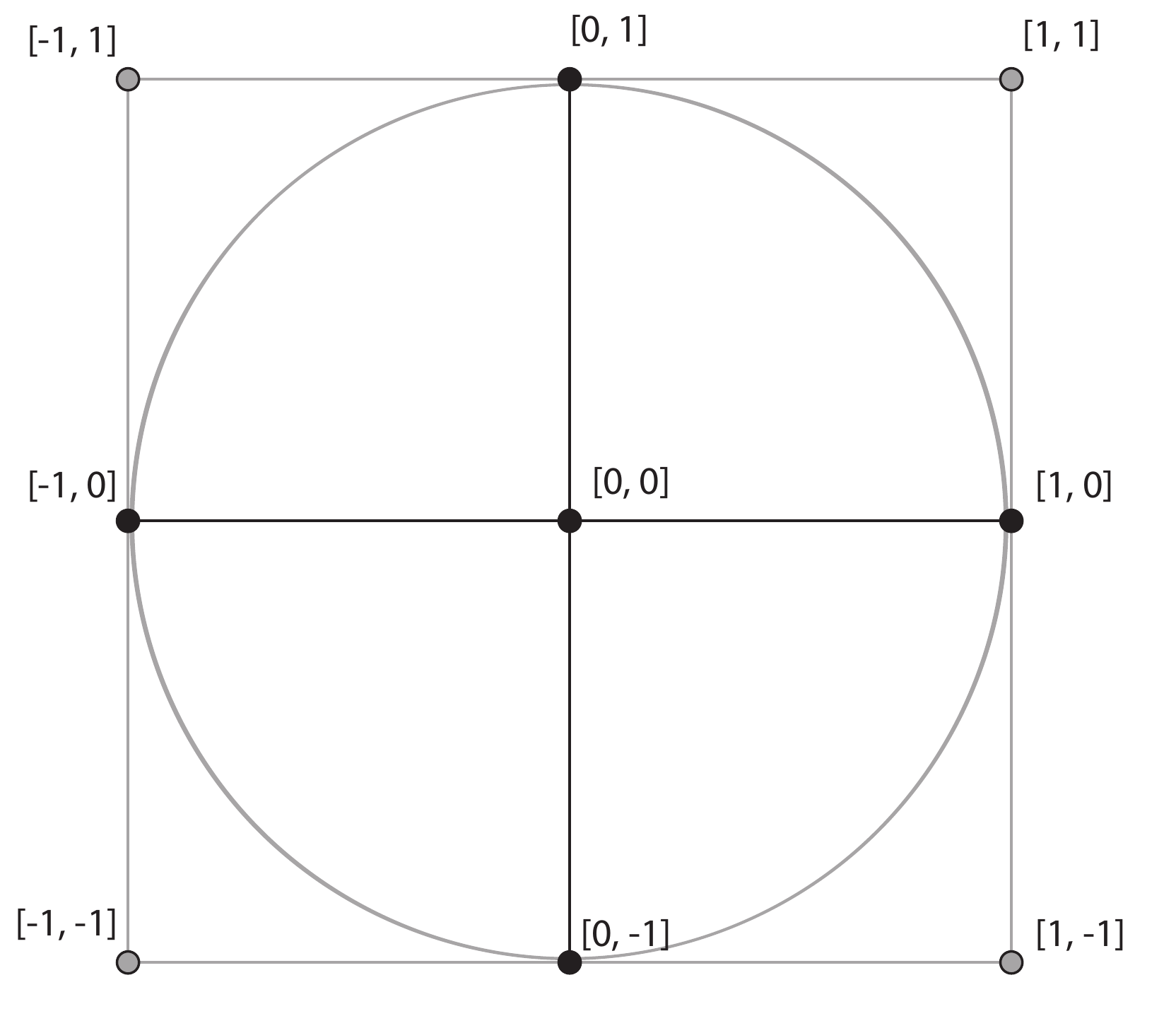}
\caption{\small \sl The one-factor-at-a-time method can only reach
  points on the cross. In this simple two-dimensional parameter space,
  each discrete factors can only take values 0, 1 or -1. OAT can reach
  $[0, 0]$, $[0, 1]$, $[0, -1]$ (points on the vertical line); and [-1, 0],
  $[1, 0]$ (points on the horizontal line). The points explored are thus
  on a cross. The points not explored are the corners $[-1, 1]$, $[-1, -1]$,
  $[1, 1]$, $[1, -1]$. In a geometrical interpretation, note that the cross
  is by construction inscribed in the circle. But OAT actually restricts
  the exploration to points on the cross, not inside the circle because
  exploring points inside the circle would imply varying two parameters
  at the same time. Now, that cross itself is inscribed in the
  circle. In sum, OAT restricts the exploration to the cross, not the
  circle, but the cross is inscribed in the circle. And this circle is
  inscribed in the square (2-cube), which is why OAT can't reach the
  corners of the square. \label{fig:figure1}}
\end{center}  
\end{figure} 

Let us now generalize this example with a geometrical interpretation of
the parameter space. In n-dimensions, the n-cross will necessarily be
inscribed in the n-sphere. The problem is that this n-sphere represents
a small percentage of the total parameter space defined by the
n-cube. This is illustrated in Figure 1, where the cross explored is
inscribed in the circle of center $[0, 0]$ and radius 1. In this
2-dimensional example, the ratio of the partially explored to the total
area --i.e. the square minus the circle-- is $r \approx 0.78$. The
problem gets quickly worse as we increase the number of dimensions. In 3
dimensions, $r \approx 0.52$ and in 12 dimensions, $r \approx
0.000326$ (see \cite{saltellipaper} for those calculations, as well as
critiques and alternatives to OAT).

Fine-tuning arguments typically vary one parameter at a time. So, they
use the OAT method to explore the space of alternative universes by
varying one by one some of the 31 fundamental physics and cosmic
parameters. They actually explore only $r \approx 4.56 \cdot 10^{-15}$
of the parameter space. We conclude that such fine-tuning arguments have
restricted their exploration to $0.00000000000000456\%$ of the relevant
parameter space!%
\footnote{\,I used the formulae in (\cite{saltellipaper}, 1510) for this
  calculation. Note that this assumes that we can put upper and lower
  boundaries on each of the parameters, which is not at all warranted
  for physics and cosmic parameters. Note also that this is a very
  generous estimate, since the actual exploration of OAT will only be a
  tiny n-cross within the volume of the n-sphere, which itself
  represents only $4.56 \cdot 10^{-15}$ of the full parameter space
  defined by the n-cube.} 
Can we hope to explore more of this space? How can we proceed?

Let us first call a \textit{fecund universe} a universe generating at
least as much complexity as our own. \textit{Are fecund universes rare
  or common in the multiverse?} This is the core issue of
fine-tuning. To answer it demands to explore this virtual
multiverse. Milan \'Cirkovi\'c \cite{cirkovic} and I both converged on
this conclusion. \'Cirkovi\'c used the metaphor of sailing the
archipelago of possible universes; I proposed to perform simulations of
possible universes, an endeavor called \textit{Artificial Cosmogenesis}
(or ACosm, see \cite{vidalacosm};\cite{vidaledu}; and also
\cite{vaas-vidala}; \cite{vaas-vidalb} for critiques; and
\cite{vidalreply} for replies). Such simulations would enable us not
only to understand our own universe (with ``real-world modelling'', or
processes-as-we-know-them) but also other \textit{possible} universes
(with ``artificial-world modelling'', or processes-as-they-could-be). We
thus need to develop methods, concepts and simulation tools to explore
the space of possible universes (the ``cosmic landscape'' as Leonard
Susskind \cite{susskind} calls it in the framework of string theory). In
\cite{vidalacosm}, I proposed to call this new field of research
{\textit{Artificial Cosmogenesis} because it sets forth a ``general
cosmology'', in analogy with Artificial Life (ALife) which
appeared with the help of computer simulations to enquiry about a
``general biology''. However, recent work on the EvoGrid}%
\footnote{\href{http://www.evogrid.org/}{http://www.evogrid.org}}
simulation project suggests that the growth of complexity is more likely
to remain open-ended if stochastic, non-deterministic processing is used
at the bottom, instead of deterministic rules, like in ALife.

Now that we have a framework to define possible universes, we will need
to generalize the Cosmic Evolution Equation we used to assess the
robustness of our universe to explore not only our universe $m^*$, but
also all universes $m$ element of the wider class of possible universes
$M$. This constitutes a rigorous approach to assess how fine-tuned our
universe is. However, it is important to understand that the results of
such studies would not \textit{ipso facto} provide an
\textit{explanation} of fine-tuning. Only if it turns out that our kind
of complex universe is common, then an explanation of fine-tuning would
be a principle of \textit{fecundity}: ``there is no fine-tuning, because
intelligent life of some form will emerge under extremely varied
circumstances'' (\cite{tegmark}, p4).

Most fine-tuning arguments change just one parameter at a time and
conclude that the resulting universe is not fit for developing
complexity. This leads to the ``one-factor-at-a-time'' paradox. What if
we would change \textit{several} parameters at the same time?
Systematically exploring the multiple variation of parameters seems like
a very cumbersome enterprise. As Gribbin and Rees wrote (\cite{gribbin},
p269):

\begin{quote}
If we modify the value of one of the fundamental constants, something
invariably goes wrong, leading to a universe that is inhospitable to
life as we know it. When we adjust a second constant in an attempt to
fix the problem(s), the result, generally, is to create three new
problems for every one that we ``solve''. The
conditions in our universe really do seem to be uniquely suitable for
life forms like ourselves, and perhaps even for any form of organic
complexity.
\end{quote}

Back in 1991, it indeed seemed very difficult to explore and find
alternative universe. However, a way to overcome this problem is to use
\textit{computer simulations} to test systematical modifications of
parameters' values. In varying just one parameter, parameter sensitivity
arguments have only begun to explore possible universes, like a baby
wetting his toes for the first time on the seashore. Surely, we had to
start somewhere. But it is truly a tiny exploration. Furthermore, maybe
there is a deep link between the different constants and physical laws,
such that it makes no sense to change just one parameter at a
time. Changing a parameter would automatically perturb other parameters
(see \cite{bradford}, p1581). Fortunately, more recent research have
gone much further than these one-parameter variations.

What happens when we vary multiple parameters? Let us first generalize
the Cosmic Evolution Equation, which this time includes other possible
cosmic \textit{evolutions} -- notice the plural! Let us imagine that we
run multiple times simulations of different models of universes $m$. We
interpret the results of the multiple runs of the simulations as a set
of \textit{virtual} universes. We end up with a distribution function
$f(m)$ combining the probability distributions obtained for each factor
of the CEE. Let us mention that, based on modern developments in
computer science, there is another more theoretical way to study and
choose distribution functions for possible universes (see the remarkable
study of Schmidhuber \cite{schmidhuber}).

The \textit{possibility space} is the huge $M$ resulting from the
definition of possible universes; and we add a measure $\pi$ on $M$. The
resulting ensemble of simulated universes $E$ would thus be defined as:
\begin{equation*}      
E=\{M, \pi, f(m)\}
\end{equation*}
The number of planets with intelligent life would then be:
\begin{equation*}      
\textit{N}_{life} (m)  = \int \textit{ N}_g \cdot  \textit{ N}_S \cdot  \textit{ f}_S \cdot \textit{ f}_p \cdot  \textit{ n}_e \cdot \textit{ f}_l \cdot \textit{ f}_i \cdot \pi
\end{equation*}
We are now talking about cosmic outcomes in other universes. The topic
becomes quite speculative, because it is not clear at all \textit{which}
cosmic outcomes are the most relevant to assess. The factors in the
equation above might be totally irrelevant. What if other possible
universes do not generate objects like galaxies, stars and planets, but
completely different kinds of complex structures? Nothing \textit{that
  we know} may evolve anymore... but other things might! We now see the
fundamental importance to define cosmic outcomes and the emergence of
complexity in a very general manner, so they can also apply to other
possible universes. Bradford \cite{bradford} proposed such a framework
when he wrote about sequences of entropy reduction. Aunger's
\cite{aungera} systems theoretical approach in terms of energy
innovation, organization and control is also a higher-level
approach. Valentin Turchin \cite{turchin} also proposed a cybernetic
theory of complexity transitions with the central concept of
\textit{metasystem transition}. Theoretical computer science measures
such as \textit{algorithmic complexity} (see e.g. \cite{li}) or
\textit{logical depth} \cite{bennett} are also precious tools to assess
the complexity of systems in a universal manner. But these are just a
few examples of frameworks to tackle the general, fascinating and
fundamental problems of the evolution and measure of complexity (see
also \cite{bedau} for a discussion in the context of Artificial Life).

We already saw that higher outcomes $f_l \cdot f_i$
are harder to assess. This is precisely where computer simulations can
be very helpful. Typically, there are so many local interactions in the
evolution of complex organisms that it is hard to analyze them
analytically with a deterministic science approach. For example, there
is not one single equation which allows to predict the development of an
embryo.

Let us now outline some remarkable alternative complex universes that
researchers recently studied. Gordon McCabe studied variations on the
standard model of particles, by changing the geometrical structure of
space-time. The result is not the end of any complexity, but just the
beginning of a new set of elementary particles. McCabe (\cite{mccabe},
2:38) elaborates:
\begin{quote}
Universes of a different dimension and/or geometrical signature, will
possess a different local symmetry group, and will therefore possess
different sets of possible elementary particles. Moreover, even
universes of the same dimension and geometrical signature will not
necessarily possess the same sets of possible particles. To reiterate,
the dimension and geometrical signature merely determines the largest
possible local symmetry group, and universes with different gauge
fields, and different couplings between the gauge fields and matter
fields, will possess different local symmetry groups, and, perforce,
will possess different sets of possible particles.
\end{quote}
It thus seems that we can vary basic physics parameters without
compromising all kinds of cosmic evolution. Who knows what kind of
complexity can emerge from this new set of particles?

As an illustration of their framework to define the multiverse, Ellis,
Kirchner and Stoeger \cite{elliskirchstoe} did examine some parameter
variations in Friedmann-Lema\^{\i}tre-Robertson-Walker (FLRW)
models. They found life-allowing regions in a phase space described by
the evolution of FLRW models. The fact that they found \textit{regions}
and not a \textit{single point }in the phase space shows that there is
room for some variation. So it seems that we can vary fundamental
geometrical cosmological parameters without precluding the apparition of
life.

Harnik, Kribs and Perez \cite{harnik} constructed a universe without
electroweak interactions called the Weakless Universe. They show that by
adjusting standard model and cosmological parameters, they are able to
obtain:

\begin{quote}
a universe that is remarkably similar to our own. This
``Weakless Universe'' has big-bang nucleosynthesis, structure formation,
star formation, stellar burning with a wide range of timescales, stellar
nucleosynthesis up to iron and slightly beyond, and mechanisms to
disperse heavy elements through type Ia supernovae and stellar mergers.
\end{quote}

This is a truly remarkable result because the cosmic outcomes are
numerous, relatively high and non trivial. Three factors in the CEE are
addressed more or less directly: $\textit{ N}_g \cdot \textit{ N}_S
\cdot \textit{ f}_S$. Maybe strong living creatures could live in the
weakless universe? This remains to be investigated.

Anthony Aguire \cite{aguirre} did study a class of cosmological models
``in which some or all of the cosmological parameters differ by orders
of magnitude from the values they assume in the standard hot big-bang
cosmology, without precluding in any obvious way the existence of
intelligent life.'' This study also shows that it is possible to vary
parameters widely without obviously harming the emergence of complexity
as we know it.

Robert Jaffe, Alejandro Jenkins and Itamar Kimchi \cite{jaffe} pursued a
detailed study of possible universes with modified quark masses. They
define \textit{congenial} worlds the ones in which the quark masses
allow organic chemistry. Again, they found comfortable regions of
congeniality.

Fred C. Adams \cite{adams} has conducted a parametric survey of stellar
stability. He found that a wide region of the parameter space provides
stellar objects with nuclear fusion. He concludes that the ``set of
parameters necessary to support stars are not particularly rare.''

An early attempt to explore alternative universes with simulations has
been proposed by Victor Stenger \cite{stengera, stengerb}. He has
performed a remarkable simulation of possible universes. He considers
four fundamental constants, the strength of electromagnetism $\alpha$;
the strong nuclear force $\alpha_s$, and the masses of the electron and
the proton. He then analysed ``100 universes in which the values of the
four parameters were generated randomly from a range five orders of
magnitude above to five orders of magnitude below their values in our
universe, that is, over a total range of ten orders of magnitude''
\cite{stengerb}. The distribution of stellar lifetimes in those
universes shows that most universes have stars that live long enough to
allow stellar evolution and heavy elements nucleosynthesis. Stenger's
initial motivation was to refute fine-tuning arguments, which is why he
ironically baptised his simulation ``MonkeyGod''. The implicit idea is
that even a stupid monkey playing with cosmic parameters can create as
much complexity as God.

In conclusion, other possible universes are also fine-tuned for some
sort of complexity! Those remarkable studies show consistently that
\textit{ alternative complex universes are possible}. One might object
that such explorations do not yet assess the higher complexity factors
in the CEE. They do not answer the following key questions: would other
interesting complex structures like planetary systems, life,
intelligence or technology evolve in those other universes? However,
these are only early attempts in conceptualizing and simulating other
possible universes, and the enterprise is certainly worth pursuing. The
fine-tuning issue could then be seriously tackled, because we would know
more and more precisely the likelihood of having our universe as it is,
by comparing it to other possible universes. Such pioneering studies are
just a beginning, and certainly new studies will come up with more and
more complex alternative universes.

\section{Summary}\label{section6}

Let us now summarize the three main steps necessary to assess how
fine-tuned our universe is.
\begin{enumerate}

\item \textit{Define} a space $M$ of possible universes
\item \textit{Explore} this space
\item \textit{Assess} the place of our universe in $M$
\end{enumerate}
Let us review step (1). Our analysis of the historical trends regarding
free parameters \cite{vidaledu} invites us to start with a \textit{weak
  variation}, i.e. varying free parameters in physical and cosmological
models. Why not vary the laws of physics themselves? It seems a very
cumbersome enterprise, because we do not even know how to make them vary
(see \cite{vaas-smolin}). It can also be dubious to do so, since the
distinction between laws and initial or boundary conditions is fuzzy in
cosmology \cite{ellisb}.

This suggestion to focus on weak variation makes most sense for the
following reasons. First, it is concrete and operational, and has a
clear meaning with well established physics. Second, we assume
supernatural miracles happening in the middle of cosmic evolution to be
-- by definition -- impossible. We assume there is a consistency and
continuity in cosmic evolution. We hypothesize that higher level
parameters are ultimately reducible to these physics and cosmic
ones. The emergent higher levels occur naturalistically. Of course, this
remains to be shown, and for practical purposes we might assume as given
such higher level parameters in our studies and simulations. New levels
of emergence, new levels of complexity did historically emerge from
lower levels, even if complicated top-down causation occurs (see
e.g. \cite{ellisc}). Take for example an economic law like the law of
supply and demand. It did not and could not exist before the apparition
of organized human civilizations. It emerged out of such new
organizations. It seems that what we call ``natural laws'' are simply
the result of more and more regular interactions. For example, as the
universe cools down, new organizations emerge. Again, it is clear that a
few billion years ago, there was no economic laws.

We also need to be more specific to apply probabilities to the ensemble
of possible universes, and avoid probabilistic fallacies. For example,
we must decide, arbitrarily or not, parameter's upper and lower
bounds. This is necessary for all practical purposes, because we can not
explore the parameter space of all parameters varying from $-{\infty}$
to $+{\infty}$. We thus need to define the maximum deviation allowed for
each parameter.

We must beware of one-factor-at-a-time limitations and paradox. We must
also define a probability measure on the parameter space. I refer the
reader to \cite{koperski} and \cite{elliskirchstoe} for detailed
arguments that measure-theoretical grounds can be specified to assess
fine-tuning. It is also crucial to define \textit{cosmic outcomes} to
specify the object of fine-tuning we aim to address. Do we talk about
fine-tuning for nucleosynthesis? atoms? Stars? Life? Intelligence? Or a
more general complexity emergence?

Step (2) requires to explore this space. The simplest exploration is to
re-run the tape of \textit{our} universe. But this only tackles the
issue of the \textit{robustness} of the universe. If we want to address
the fine-tuning issue we must also run and re-run tapes of \textit{other
  possible universes}. This will bring us insights into how our and
other universes are parameter sensitive, and generate complex
outcomes. Although we always need good theoretical models to start with,
it is necessary to use computer simulations to explore the huge
parameter landscape we are talking about. That landscape is not just
very big, but really huge. Because we don't want to and do not have the
resources to explore the space blindly, it also makes most sense to use
simulations to test particular hypotheses and theories. As an
application, if we take Lee Smolin's \cite{smolin} cosmological natural
selection theory, and find alternative universes with more black holes
(the cosmic outcome under consideration) by tweaking parameters, it is a
way to falsify the theory.

The last step (3) is to compare the distribution functions of the cosmic
outcomes obtained through simulations, to the space $M$ of possible
universes. In other words, we assess the probability to find a universe
with outcome O. Note that this is the crucial difference between
tackling the robustness and the fine-tuning issue. In robustness
analysis, we run multiple times the \textit{same} universe simulation
changing only the random dynamical parameters. We compare multiple runs
of the same universe. In fine-tuning analysis, we run multiple
\textit{different} universe simulations, changing a wide number of
parameters. We compare our universe to the set of possible
universes. How typical or atypical is our universe in the space of
possible universes? The results of such simulation experiments will
enable us to answer this question. Ideally, we will be in a position to
assess the likelihood or unlikelihood of complexity emergence in the
space of possible universes. Even better than assessing specific cosmic
outcomes, which might bias us to a universe-centric perspective, we can
aim to assess the probability to find universes which display open-ended
evolutionary mechanisms leading to ever increasingly complex cosmic
outcomes.

To the traditionally trained cosmologist, this enterprise might seem
totally unconventional. And it is, because it is a new kind of
computational science. This is why we can call it \textit{Artificial
  Cosmogenesis}. It might also seem out of reach. As I argued elsewhere,
since the sheer computational resources grow more than exponentially,
this allows us in principle to increase accordingly the complexity and
richness of our computer simulations \cite{vidalacosm}.  Additionally,
engineers and professional model makers have developed a wide variety of
tools to test multiple variables, rarely used in cosmological
contexts. Let us just mention of few of them. A starting point is to use
the tools of global sensitivity analysis (see e.g. \cite{saltellibook}).
These include advanced statistical approaches such as latin hypercube
sampling, multivariate stratified sampling or Montecarlo simulations for
finding dynamic confidence intervals. Systems dynamics and engineering
have also many tools to offer such as phase portraits or probabilistic
designs. The classic book by John D. Sterman \cite{sterman} remains a
reference and quite comprehensive introductory book on complex systems
modeling and simulations.

Let us now be scrupulous. What is a proof of fine-tuning? Let $n$ be the
number of free parameters. We have a logical and statistical version of
what a proof of fine-tuning would be:
\pagebreak

\begin{quote}
\textbf{Logical proof of fine-tuning: }\textit{If you vary one
  parameter, there exists no possible universe generating outcome $O$ by
  adjusting the $(n-1)$ other parameters.} \newline Which is equivalent
to:\newline \textit{if you vary one parameter, there is no way
  whatsoever that any other possible universe can generate outcome
  $O$.}
\end{quote}

\begin{quote}
\textbf{Probabilistic proof of fine-tuning: } \textit{If you vary one
  parameter, adjusting the $(n-1)$ other parameters will not make
  outcome $O$ more likely.} \newline Which is equivalent to:
\newline\textit{if you vary one parameter, there is no way whatsoever
  that any other possible universe can generate outcome $O$ with a
  higher probability.}
\end{quote}
\bigskip

\noindent In sum, you need to have explored the relevant parameter space
of possible universes to make serious claims about fine-tuning. Pretty
hard to prove! This is even harder for outcomes as advanced as life or
intelligence.  Our conclusion is that \textit{fine-tuning for life or
  intelligence remains a conjecture}. Like in mathematics, we have
strong reasons to believe the conjecture is true, but a proof is out of
reach and certainly requires a huge amount of work. As a matter of fact,
the challenge of simulating possible universes and comparing them is
overwhelming. This is why the concept of the cosmic outcome is so
important to ease the process. Indeed, we can break down the problem and
progress by tackling higher and higher outcomes, with more and more
connection between outcomes. We don't need nor can assess all outcomes
at once in the CEE. As our understanding, modeling capacities and
computational resources increase, we can be more ambitious in simulating
more and more as well as higher and higher outcomes in cosmic
evolution. I am well aware of the highly ambitious research program that
ACosm proposes. However, the good news is that there is work for many
generations of scientists. Tomorrow's cosmology is not restricted to
empirical observations or highly theoretical models. It is also the
science of simulating and experimenting with alternative universes.

\section{Conclusion}
Up to now, discussions about possible universes were chiefly a
metaphysical recreation. We advanced conceptual foundations to study
possible universes scientifically, with the help of computer
simulations. This approach is needed if we take seriously the thesis of
computational irreducibility, namely that most complex systems are
theoretically impossible to predict in a deterministic or statistical
manner. A more general computational kind of science is needed. We
applied this new kind of science to cosmology, to address two key
cosmological issues: the robustness of the emergence of complexity, and
the fine-tuning of the universe.

We first formulated the issues of defining possible universes, and
possible cosmic outcomes (sections \ref{section2} and \ref{section3}).
Based on previous work, we defined a modular ``Cosmic Evolution
Equation'' (CEE). This equation can have many applications to define
research agendas in computational cosmology. In particular, to tackle
our two issues, we adjusted the CEE by varying the space of possible
universes it acts upon, to study either the robustness (section
\ref{section4}) or the fine-tuning issue (\ref{section5}).

Importantly, we considered only a \textit{virtual multiverse}, that we
define within our concrete models and simulations. This is in sharp
contrast with speculations about an actual multiverse, an idea quite
common in modern cosmology, yet often criticized for being hard or
impossible to test scientifically.

To address the delicate fine-tuning issue, we further argued that
studies and simulations of alternative possible universes are demanded,
a research field called \textit{Artificial Cosmogenesis} (ACosm,
sections \ref{section5}-\ref{section6}). This field is actually not new,
since we outlined quite some research which have examined alternative
possible universes. Yet these studies are really just beginning to
explore possible universes, and ACosm holds great promise to further
investigate whether and how our universe and others generate increasing
complexity.

\section{Acknowledgments}

I thank Rick Bradford, Bruce Damer, Jean-Paul Delahaye, Francis
Heylighen, Tom\'as Igor Veloz Gonz\'alez, Tobias Kerzenmacher, Stanley Salthe and William Stoeger for thoughtful comments and criticisms.

\section{Appendix - Argumentative Maps}\label{maps}

Figure 2. maps the problem described in introduction, while Figure
3. maps the core argument presented in the paper. Please read in a
top-down direction. More details on argumentation mapping can be found
in \cite{vidalacosm}.

\begin{figure} 
\centering\includegraphics[clip=true, viewport=60 0 765 650,width=14.5cm]{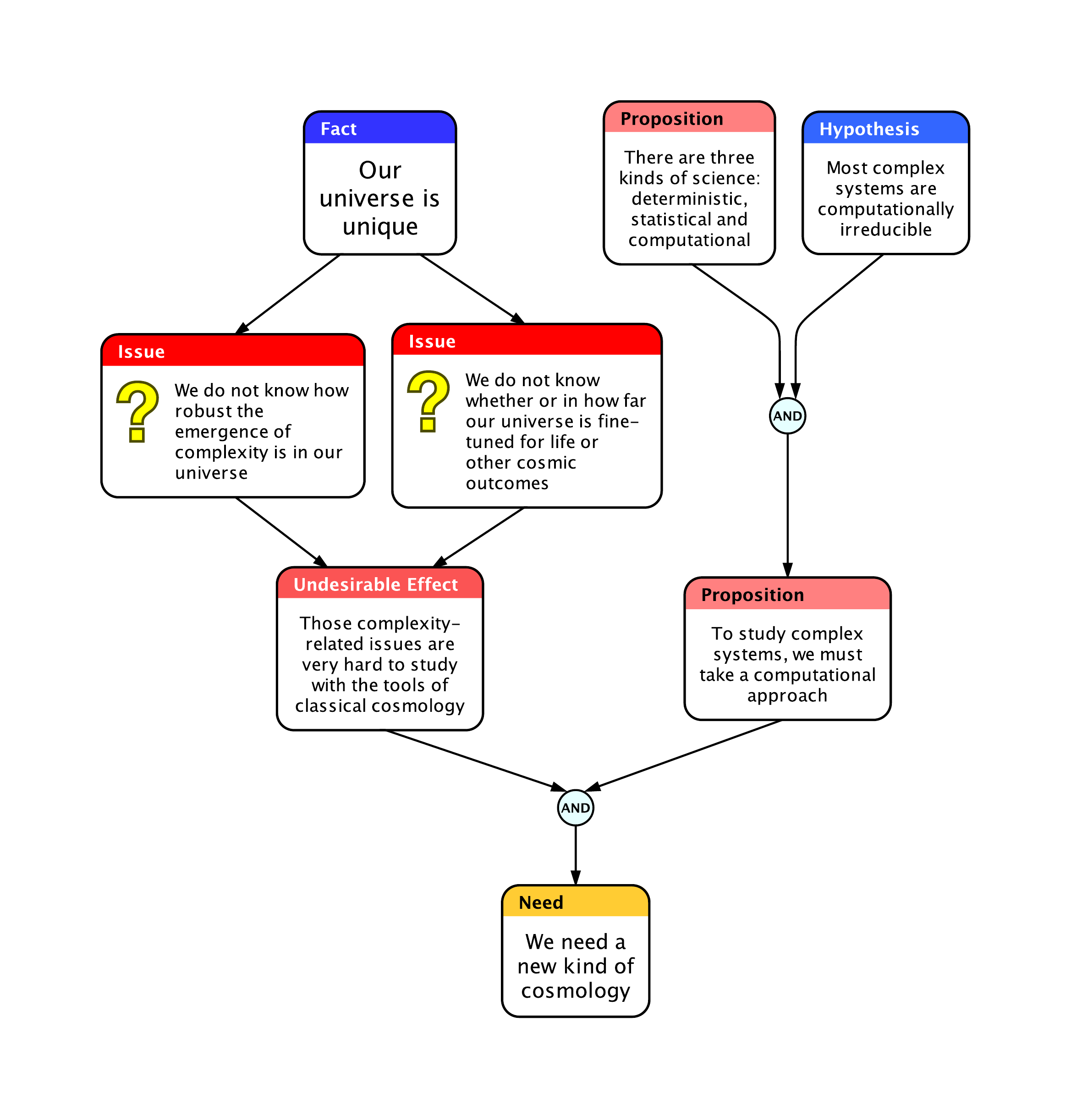}
\caption{\small \sl The core problem \label{fig:figure2}} 
\end{figure}

\begin{figure} 
\begin{center} 
\includegraphics[width=9.128cm,height=15.981cm]{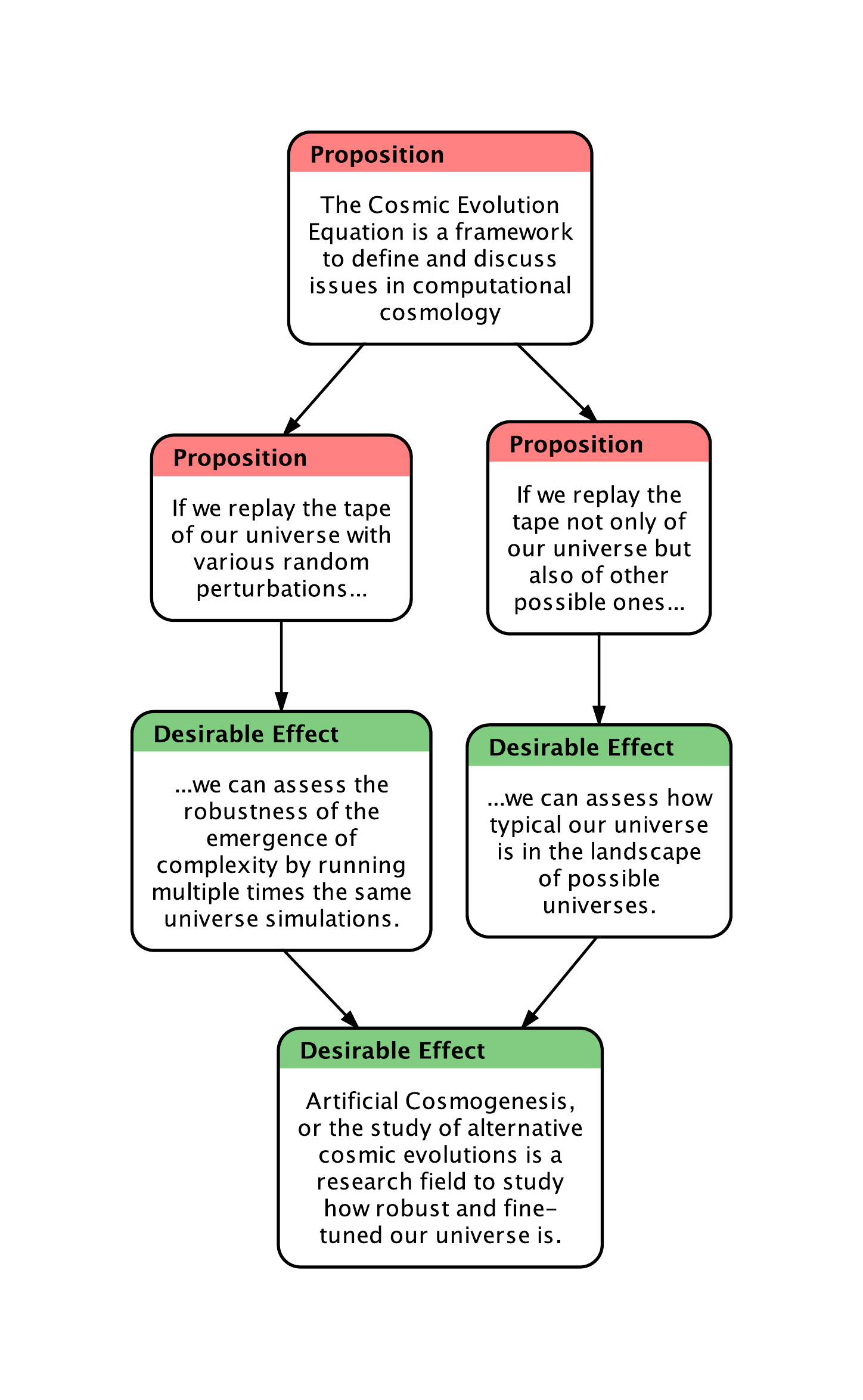}
\caption{\small \sl The proposed solution \label{fig:figure3}} 
\end{center}  
\end{figure}

\pagebreak

\end{document}